%% file: paper_jcgs.tex
\documentclass[letterpaper,12pt]{article}

\usepackage[margin=1in]{geometry}
\usepackage{charter}
\usepackage{amsmath,amsthm,amsfonts,amssymb,bm}
\usepackage{algorithm,algorithmic}
\usepackage{enumerate}
\usepackage{epsfig,rotating,color}
\usepackage{paralist,xspace,setspace,natbib}
\usepackage{tabularx,booktabs}
\usepackage{hyperref}
 \rm

\usepackage{xr}
\hypersetup{colorlinks}
\hypersetup{citecolor = black}
\hypersetup{linkcolor = black}
\hypersetup{urlcolor  = black}
\hypersetup{filecolor = black}

\input{macros}

\numberwithin{equation}{section}
\numberwithin{corollary}{section}
\numberwithin{lemma}{section}
\numberwithin{theorem}{section}

\title{Towards Automatic Model Comparison:\\An Adaptive Sequential Monte Carlo
  Approach}
\author{Yan Zhou{$\vphantom{u}^1$}, Adam M. Johansen{$\vphantom{n}^2$} and
  John A. D. Aston{$\vphantom{n}^3$} }
\begin{document}

\maketitle

\abstract {%
  \singlespace\noindent
  Model comparison for the purposes of selection, averaging and validation is
  a problem found throughout statistics. Within the Bayesian paradigm, these
  problems all require the calculation of the posterior probabilities of
  models within a particular class. Substantial progress has been made in
  recent years, but difficulties remain in the implementation of existing
  schemes. This paper presents adaptive sequential Monte Carlo (\smc) sampling
  strategies to characterise the posterior distribution of a collection of
  models, as well as the parameters of those models. Both a simple product
  estimator and a combination of \smc and a path sampling estimator are
  considered and existing theoretical results are extended to include the path
  sampling variant. A novel approach to the automatic specification of
  distributions within \smc algorithms is presented and shown to outperform
  the state of the art in this area. The performance of the proposed
  strategies is demonstrated via an extensive empirical study. Comparisons
  with state of the art algorithms show that the proposed algorithms are
  always competitive, and often substantially superior to alternative
  techniques, at equal computational cost and considerably less
  application-specific implementation effort.

}

\section{Introduction}
\label{sec:Introduction}

Model comparison lies at the core of Bayesian decision
theory \citep{Robert:2007tc} and has attracted considerable attention in recent
decades. Most approaches to the calculation of the required
posterior model probabilities depend upon asymptotic arguments,
the post-processing of outputs from Markov chain Monte Carlo (\mcmc)
algorithms operating on the space of a single model or using specially
designed \mcmc techniques that provide direct estimates of these quantities
(e.g. Reversible Jump \mcmc, \rjmcmc; \citet{Green:1995dg}).
Within-model simulations are simpler, but generalisations
of the harmonic mean estimator \citep{Gelfand:1994ux} which are widely used
in this setting require careful design to ensure finite variances and,
convergence assessment can be difficult. Simulations on the whole model
spaces are often difficult to implement efficiently even though they can be
conceptually appealing.

More robust and efficient Monte Carlo algorithms have been established in
recent years. Many of them are population based, dealing with a collection of
samples at each iteration, including sequential importance sampling and
resampling (Annealed Importance Sampling \ais, \citet{Neal:2001we}; Sequential
Monte Carlo \smc, \citep{DelMoral:2006hc}) and population \mcmc (\pmcmc;
\citet{Liang:2001dc,Jasra:2007in}). However, most studies have focused on
their abilities to explore high dimensional and multimodal spaces. The
application of these algorithms to Bayesian model comparison is less well
studied. Here, we motivate and present approaches based around the \smc family
of algorithms, and demonstrate their effectiveness empirically.

\smc methods are a class of sampling algorithms which
combine importance sampling and resampling. They have been primarily used as
``particle filters'' to solve optimal filtering problems; see, for example,
\citet{Cappe:2007hz,Doucet:2011us} for recent reviews. They are used here in a
different manner, that proposed by \citet{DelMoral:2006hc} and developed by
\citet{DelMoral:2006wv,Peters:2005wh}. This framework employs a sequence of
artificial distributions on spaces of increasing dimensions which admit the
distributions of interest as marginals.

% Resampling is stochastic elimination of samples with small weights and
% replication of those with large weights to produce an unweighted sample
% suitable for approximating the same distribution as the original weighted
% sample; resampling concentrates computational resources appropriately
% without introducing bias.

% Application of standard sequential importance resampling algorithms to these
% extended distributions provides samples suitable for the approximation of
% each distribution of interest in turn. As is usual in the \smc literature,
% we advocate the use of low-variance resampling schemes \citet{Douc:2005wa};
% resampling adaptively according to an appropriate criterion such as
% effective sample size (\ess; \citet{DelMoral:2012jq}). All distributions are
% assumed to admit densities with respect to some dominating measure solely to
% lighten notation.

Although it is well known that \smc is well suited to the computation of
normalising constants and that it is possible to develop relatively automatic
\smc algorithms by employing a variety of ``adaptive'' strategies, their use
for Bayesian model comparison has not yet received a great deal of attention.
We highlight three strategies for computing posterior model probabilities
using \smc, focusing on strategies which require minimal tuning and can be
readily implemented requiring only the availability of \emph{locally-mixing}
\mcmc proposals. These methods admit natural and scalable parallelisation and
we demonstrate the potential of these algorithms with real implementations
suitable for use on consumer-grade parallel computing hardware including
\gpu{}s, reinforcing the message of \citet{Lee:2010fm}. We also present a new
approach to adaptation and guidelines on the near-automatic implementation of
the proposed algorithms. These techniques are applicable to \smc algorithms in
much greater generality. The proposed approach is compared with
state of the art alternatives in extensive simulation studies which
demonstrate its performance and robustness.

The next section we provides a brief survey of Bayesian model comparison literature.
Section~\ref{sec:Methodology} presents three algorithms for performing model
comparison using \smc techniques and Section~\ref{sec:Illustrative
  Applications} provides several illustrative applications, together with
comparisons with other techniques. The paper concludes with some discussion.
%s in Section~\ref{sec:Conclusion}.
%The appendix gives more details on the method for adaptively choosing the distributions. In addition, the supplementary material gives a central limit theorem for the path sampling estimators in the suggested methodology.

\section{Background}
\label{sec:Background}

Bayesian model comparison depends upon the posterior distribution over
models. It is only possible to obtain closed-form expressions for posterior
model probabilities in very limited situations.
The general problem has attracted considerable attention and it is not feasible to
exhaustively summarise this literature here. We describe %aim only to describe the major
the major contributions to the area and recent developments of particular relevance.

\subsection{Analytic Methods and MCMC}
\label{sub:Conventional Methods}

The Bayesian Information Criterion (\bic), developed by
\citet{Schwarz:1978uv}, is based upon a large sample approximation of the
Bayes factor.
% It is defined as $\text{\bic} = -2\widehat\ell + k\log(n)$, where
% $\widehat\ell$ denotes the maximum of the log-likelihood for the observed
% data, $k$ the number of model parameters and $n$ the effective dimension of
% the data.
An
asymptotic argument concerning Bayes factors under appropriate regularity
conditions justifies the choice of the model with the smallest value of \bic.
Although appealing in its simplicity,
%such an approach can only be formally
%justified when a
justification requires the availability of a large number of observations.% (compared to the
%number of parameters) is available.

%In this era of fast computing, the difficulty of evaluating the integrals that
%must be computed in order to adapt a fully Bayesian approach is much smaller
%than it once was.
The Bayesian approach to model comparison is, of course, to
consider the posterior probabilities of the possible models \citep[Chapter
6]{Bernardo:1994vd}.
%Within this Bayesian framework the decision making
%process, which might include model comparison, model selection or the choice
%of an action, depend upon the relative probabilities of several models
%\citep[Chapter 7]{Robert:2007tc}.

Given a denumerable collection of models $\{\Mk\}_{k\in\mset}$, with
model $\Mk$ having parameter space $\Paramk$, Bayesian inference proceeds from
a prior distribution over the collection of models, $\pi(\Mk)$, a prior
distribution for the parameters of each model, $\pi(\paramk|\Mk)$ and the
(model-specific) likelihood $p(\data|\paramk,\Mk)$ to the model posterior:
\begin{equation}
  \pi(\Mk|\data) = \frac{p(\data|\Mk)\pi(\Mk)}{p(\data)},
\end{equation}
where $p(\data|\Mk) = \int_{\paramk} p(\data|\paramk,\Mk) \pi(\paramk|\Mk)
\intd\paramk$ is termed the \emph{evidence} for model $\Mk$ and the
normalising constant $p(\data) = \sum_{k\in\mset}p(\data|\Mk) \pi(\Mk)$ can be
easily calculated if $|\mset|$ is finite and the evidence for each model is
available. The case where $|\mset|$ is countable is discussed later. We first
review some techniques for evidence  calculation.
% the evidence for each model individually.

Several techniques have been proposed to approximate the evidence for a model
using simulation techniques which approximate the posterior distribution of
that model, including the harmonic mean estimator of
\citet{Newton:1994wm,Raftery:2007ud} and generalisations thereof
\citet{Gelfand:1994ux}. These pseudo-harmonic mean methods use the insight
that for any density $g$, such that $g(\cdot) \ll p(\cdot|\data,\Mk)$, the following identity holds,
\begin{equation}
  \int \frac{g(\paramk)}{p(\data,\paramk|\Mk)}
  \pi(\paramk|\data,\Mk) \intd\paramk
  = \int \frac{g(\paramk)}{p(\data,\paramk|\Mk)}
  \frac{p(\data,\paramk|\Mk)}{p(\data|\Mk)} \intd\paramk
  = \frac{1}{p(\data|\Mk)}
\end{equation}
and by approximating the leftmost integral one can obtain an estimate of the evidence.
Unfortunately, considerable care is required in the implementation of such
schemes in order to control the variance of the resulting estimator% (and
%indeed, to ensure that this variance is finite;
--- see \citet{Neal:1994}).

In the particular case of the Gibbs sampler, \citet{Chib:1995em} provides an
alternative approach %to the approximation of the evidence from within-model
%simulations
based on the identity,
\begin{equation}
  p(\data|\Mk) =
  \frac{p(\data|\paramk,\Mk)\pi(\paramk|\Mk)}{\pi(\paramk|\data,\Mk)},
\end{equation}
which holds for any value of $\paramk$. An estimator of the marginal
likelihood can be obtained by replacing $\paramk$ with a particular value, say
$\paramk^\star$, which is usually chosen from the high probability region of
the posterior distribution and approximating the denominator
$\pi(\paramk^\star|\data,\Mk)$ using the output from a Gibbs sampler. Though
this method does not suffer the instability associated with generalised
harmonic mean estimators, it requires that all full conditional densities are
known (including their normalising constants) and that the Gibbs sampler mixes
adequately. This approach was generalised to other Metropolis-Hastings
algorithms, by \citet{Chib:2001gq}, who require only that the proposal
distributions be known.
%, where only the
%proposal distributions are required to be known.% including their normalising constants.

%The first \mcmc method which operated simultaneously over the full collection
%of models of interest providing direct estimates of posterior model
%probabilities was probably the approach of \citet{Grenander:1994vy}. However,
%
The \rjmcmc strategy first proposed by \citet{Green:1995dg} is undoubtedly the
most widespread approach that targets the joint posterior distribution over
model and parameters. \rjmcmc adapts the Metropolis-Hastings algorithm to construct a
Markov chain on an extended state-space which admits the posterior
distribution over both model and parameters as its invariant distribution.
The design of efficient between-model moves is often difficult, and the mixing
of these moves largely determines the performance of the algorithm. For
example, in multimodal models, where \rjmcmc has attracted substantial
attention, information available in the posterior distribution of a model of
any given dimension does not characterize modes that exist only in models of
higher dimension, and thus successful moves between those models become
unlikely and difficult to construct \citep{Jasra:2007id}. In addition, \rjmcmc
will not characterise models of low posterior probability well, as those
models will be visited by the chain only rarely. In some cases it will be
difficult to determine whether the low acceptance rates of between-model moves
result from actual characteristics of the posterior or from a poorly-adapted
proposal kernel.

% The related continuous time birth and death algorithm of
% \citet{Stephens:2000wq} was shown by \citet{Cappe:2003ek} to have no
% qualitative advantage over the simpler discrete time formulation.
A post-processing approach to improve the computation of normalising constants
from \rjmcmc output using a bridge-sampling approach was advocated by
\citet{Bartolucci:2006cb}. Sophisticated variants of these algorithms, such as
those developed in \citet{Peters:2010vk}, have also been considered
but depend upon essentially the same construction and ultimately require
adequate mixing of the underlying Markov process.

\citet{Carlin:1995uy} presented an alternative method for simulating the model
probability directly through a Gibbs sampler on the space $\{\Mk\}_{k\in\mset}
\times \prod_{k\in\mset}\Paramk$. The joint parameter is thus $(M,\theta)$
where $\theta$ is the vector $(\paramk)_{k\in\mset}$ and conditional on model
$\Mk$ the data $\data$ only depends on a subset, $\paramk$, of the parameters.
To form the Gibbs sampler, a so called pseudoprior $\pi(\paramk|M\ne\Mk)$ in
addition to the usual prior $\pi(\paramk|\Mk)$ is selected, such that given
the model indicator $M$, the parameters associated with different models are
conditionally mutually independent. In this way, a Gibbs sampler can be
constructed provided that all the full conditional distributions
$\pi(\paramk|\data,\paramk[k^\prime\ne k],M)$ and $\pi(M = \Mk|\data,\theta)$
for $k\in\mset$ are available. %The major drawback of this approach is that
The performance of this sampler, which was generalised by \citet{Godsill:2001cv}, is very sensitive to the selected
pseudopriors and sampling frm the full conditional
distribution must be feasible.

The methods reviewed above either demand substantial knowledge of the target
distributions or require substantial tuning.

\subsection{Recent Developments on Population-Based Methods}
\label{sub:Recent Developments on Population-Based Methods}

We consider two broad groups of population-based Monte Carlo methods.
One family, including \smc, is based on sequential importance sampling and resampling,
Another approach is population \mcmc (\pmcmc;
\citet{Marinari:1992,Geyer:1991,Liang:2001dc}) also known as parallel
tempering. \pmcmc operates by constructing a sequence of
distributions $\{\pi_t\}_{t=0}^T$ with $\pi_0$ corresponding to the target
distribution and successive elements of this sequence consisting of
distributions from which it is increasingly easy to sample. A population of
samples is maintained, with the $i^{\text{th}}$ element of the population
being approximately distributed according to $\pi_i$; the algorithm proceeds
by simulating an ensemble of parallel \mcmc chains each targeting one of these
distributions. The chains interact with one another via exchange moves, in
which the state of two adjacent chains is swapped, and this mechanism allows
for information to be propagated between the chains and hopefully for the fast
mixing of $\pi_T$ to be partially transferred to the chain associated with
$\pi_0$. The resulting samples target the product $\prod_{t=0}^T\pi_t$ which admits
$\pi_0$ as a marginal.

There is substantial interest in the use of population based methods to
explore high dimensional and multimodal parameter spaces which
challenge conventional \mcmc algorithms. \citet{Jasra:2007in} compared the
performance of the two approaches in this context. There is also increasing
interest in using these methods for Bayesian model comparison. In principle, \pmcmc
output can be post-processed in the same way as conventional \mcmc to obtain
estimates of evidence for each model. However, this approach inherits many of
the disadvantages of the basic estimators. \citet{Jasra:2007id} combined
\pmcmc with \rjmcmc and thus provide a direct estimate of the posterior model
probability. Another approach is to use the outputs from all the chains to
approximate the path sampling estimator \citep{Gelman:1998ei}, see
\citet{Calderhead:2009bd}. However, the mixing speed of \pmcmc is sensitive to
the number and placement of the distributions $\{\pi_t\}_{t=0}^T$ (see \citet{Atchade:2010ha} for the optimal placement of
distributions in terms of a particular mixing criterion for a restricted class
of models). As seen in \citet{Calderhead:2009bd}, the placement of
distributions can be critical --- see Section~\ref{sec:Illustrative
  Applications}.

The use of \ais for computing normalising constants directly and via path
sampling dates back at least to \citet{Neal:2001we}; see
\citet{Vyshemirsky:2008ch} for a recent example of its use in the computation
of model evidences. It has often been suggested that more general \smc
strategies provide no advantage over \ais when the normalizing constant is the
object of inference. Later we will demonstrate
that this is not generally true, adding improved robustness of normalizing
constant estimates to the advantages afforded by resampling within \smc. This
is consistent with theoretical results \citep{Schweizer:2012} obtained in a
slightly different context which show that resampling can qualitatively
improve the theoretical behaviour of the estimator when the initial and final
distributions differ substantially. More details on the use of \smc and path sampling for Bayesian
model selection are provided in the next section. The use of \pmcmc coupled
with path sampling was discussed in \citet{Vyshemirsky:2008ch}.

\citet{Jasra:2008bb} developed a method using a system of interacting \smc
samplers for trans-dimensional simulation. The targeting distribution $\pi$
and its space $S$ are the same as in \rjmcmc. As usual in \smc, a sequence of
distributions $\{\widetilde\pi_t\}_{t=0}^T$ with increasing dimensions are
constructed such that $\widetilde\pi_T$ admits $\pi$ as a marginal. The
algorithm starts with a set of \smc samplers with equal number of particles;
each of them targets $\widetilde\pi_{i,t}(x) \propto
\widetilde\pi_t(x)\bbI(x\in S_{i})$ up to a predefined time index $t^\star$,
such that $\{S_{i}\}$ is a partition of $S$. At time $t^\star$ particles from all
samplers are allowed to coalesce, and from this time on, all of them are
iterated with the same Markov kernel (on $S$) until the single
sampler reaches the target $\pi$.
%Each individual sampler explores only a
%portion of the parameter space and by using the information which each
%sampler gains about that region of the parameter space, with a properly
%chosen $t^\star$, the resulting sampler will be able to explore the whole
%space efficiently.
One of the three algorithms detailed in the next section coincides,
essentially, with the final stage of the approach of \citet{Jasra:2008bb}; the
other algorithms which are developed rely on a quite different strategy. We
note that subsequent to the completion of the first version of this
manuscript, a related strategy has been proposed by \citet{Karagiannis2013aisrj}. they
combine \smc and \mcmc via the mechanism of particle \mcmc
\citep{ADH10} using an \smc algorithm as a \rjmcmc
proposal. This strategy is likely to lead to better mixing than conventional
\rjmcmc algorithm but comes at considerable computational cost.

A proof-of-concept study in which several \smc approaches to the problem were
outlined was provided by \citet{Zhou:2012uz} and these approaches are
developed below. These strategies based around various combinations of path
sampling \citep{Gelman:1998ei} and \smc (as used by \citet{Johansen:2006wm} in
a rare events context and by \citet{Rousset:2006kq} in the context of the
estimation of free energy differences) or the unbiased estimation of the
normalizing constant via standard \smc techniques
\citep{DelMoral:1996,DelMoral:2006hc}.

%A number of other recent developments should be mentioned for completeness,but are not directly relevant to the problems considered here.
A strategy for
\smc-based variable selection was developed by \citet{Schafer:2011bx}; however, this
approach depends upon the precise structure of this particular problem and
does not involve the explicit computation of normalizing constants.
%Several approaches to the problem of Bayesian model
%comparison in settings in which the likelihood cannot be evaluated have been
%proposed
%\citep{Grelaud:2009gc,Didelot:2011wo,Robert:2011vx}.
%IF THIS IS OUTSIDE THE SCOPE OF THIS PAPER WHY SAY SO MUCH ABOUT IT?
%Approximated Bayesian computation (\abc) represents the almost the unique
%solution to the situation where likelihood is not available. However as shown
%in \citet{Robert:2011uv} the \abc algorithm lacks confidence when it comes to
%model comparison problems due to the fact that the algorithm involves an
%unknown loss of information induced by the use insufficient summary
%statistics. They also showed that additional empirical verification of the
%performance is necessary for model comparison purposes as the computational
%effort may not always reduce the approximation error.
%This class of problems falls outside the scope of the current paper. We will
%but we assume throughout that the likelihood of all models can be evaluated.

\subsection{Challenges for Model Comparison Techniques}
\label{sub:Difficulties of existing methods}
There are a number of
desirable features in algorithms which seek to address any model comparison
problem and that these desiderata can find themselves in competition with one
another. One always requires accurate evaluation of Bayes factors or model
proportions and to obtain these one requires estimates of either normalizing
constants or posterior model probabilities with small error making the
efficiency of any Monte Carlo algorithm employed in their estimation critical.
If one is interested in characterising behaviour conditional upon a given
model or even calculating posterior-predictive quantities, it is likely to be
necessary to explore the full parameter space of each model; this can be
difficult if one employs between-model strategies which spend little time in
models of low probability. In many settings end-users seek to interpret the
findings of model selection experiments and in such cases, accurate
characterisation of all models including those of relatively small probability
may be important.

\section{Methodology}
\label{sec:Methodology}

\smc samplers provide, iteratively, collections of weighted samples
from a sequence of distributions $\{\pi_t\}_{t=0}^T$ over essentially any
random variables on some measurable spaces $(E_t,\calE_t)$, by constructing a
sequence of auxiliary distributions $\{\widetilde\pi_t\}_{t=0}^T$ on spaces of
increasing dimensions,
\begin{equation}\label{eq:pitilde}
  \widetilde\pi_t(x_{0:t})=\pi_t(x_t)\prod_{s=0}^{t-1}L_s(x_{s+1},x_s),
\end{equation}
where the sequence of Markov kernels $\{L_s\}_{s=0}^{t-1}$, termed backward
kernels, is formally arbitrary but critically influences the estimator
variance. See \citet{DelMoral:2006hc} for further details and guidance on the
selection of these kernels.

Standard sequential importance resampling algorithms can then be applied to
the sequence of synthetic distributions, $\{\widetilde\pi_t\}_{t=0}^T$. At
time $t = n - 1$, assume that a set of weighted particles
$\{W_{n-1}^{(i)},X_{0:n-1}^{(i)}\}_{i=1}^N$ approximating
$\widetilde\pi_{n-1}$ is available, then at time $t = n$, the path of each
particle is extended with a Markov kernel say, $K_n(x_{n-1}, x_n)$ yielding the set
of particles $\{X_{0:n}^{(i)}\}_{i=1}^N$ and
% reach the distribution
%$\eta_n(x_{0:n}) = \eta_0(x_0)\prod_{t=1}^nK_t(x_{t-1}, x_t)$, where $\eta_0$
%is the initial distribution of the particles.
%To correct the discrepancy between $\eta_n$ and $\widetilde\pi_n$,
importance sampling is then applied.
The weights are update by a factor %$\widetilde{w}_n(x_{n-1},x_n)$, where
% \begin{equation}
%   W_n(x_{0:n}) \propto
%   \frac{\widetilde\pi_n(x_{0:n})}{\eta_n(x_{0:n})}
%   = \frac{\pi_n(x_n)\prod_{s=0}^{n-1}L_s(x_{s+1}, x_s)}
%   {\eta_0(x_0)\prod_{t=1}^nK_t(x_{t-1},x_t)}
%   \propto W_{n-1}(x_{0:n-1})\widetilde{w}_n(x_{n-1}, x_n)
% \end{equation}
$\widetilde{w}_n$, termed the \emph{incremental weights}, calculated
as,
\begin{equation}
  \widetilde{w}_n(x_{n-1},x_n) =
  \frac{\pi_n(x_n)L_{n-1}(x_n, x_{n-1})}{\pi_{n-1}(x_{n-1})K_n(x_{n-1}, x_n)}.
\end{equation}
If $\pi_n$ is only known up to a normalizing constant, say $\pi_n(x_n) =
\gamma_n(x_n)/Z_n$, then we can use the \emph{unnormalised} incremental
weights
\begin{equation}
  w_n(x_{n-1},x_n) =
  \frac{\gamma_n(x_n)L_{n-1}(x_n, x_{n-1})}
  {\gamma_{n-1}(x_{n-1})K_n(x_{n-1}, x_n)}
\end{equation}
for importance sampling. Further, with the previously \emph{normalised}
weights $\{W_{n-1}^{(i)}\}_{i=1}^N$, we can estimate the ratio of normalizing
constant $Z_n/Z_{n-1}$ by
\begin{equation}
  \frac{\widehat{Z}_n}{Z_{n-1}} =
  \sum_{i=1}^N W_{n-1}^{(i)}w_n(X_{n-1:n}^{(i)}),
\end{equation}
and
\begin{equation}
  \frac{\widehat{Z}_n}{Z_{1}} =
  \prod\limits_{p=2}^n \frac{\widehat{Z}_p}{Z_{p-1}} =
  \prod\limits_{p=2}^n \sum_{i=1}^N W_{p-1}^{(i)}w_p(X_{p-1:p}^{(i)}),
\end{equation}
provides an unbiased \citep[Proposition 7.4.1]{DelMoral:2004ux} estimate of $Z_n / Z_1$.
%Sequentially, the normalizing constant between initial distribution $\pi_0$
%and target $\pi_T$ can be estimated.
See \citet{DelMoral:2006hc} for details
on calculating the incremental weights in general; in practice, when $K_n$ is
$\pi_n$-invariant, $\pi_n \ll \pi_{n-1}$, and $L_{n-1}$ is the associated
time-reversal kernel,
%\begin{equation}
%  L_{n-1}(x_n, x_{n-1}) = \frac{\pi_n(x_{n-1})K_n(x_{n-1}, x_n)}{\pi_n(x_n)}
%\end{equation}
%is used as the backward kernel,
the unnormalised incremental weight function becomes
\begin{equation}
  w_n(x_{n-1},x_n) = \frac{\gamma_n(x_{n-1})}{\gamma_{n-1}(x_{n-1})}.
  \label{eq:inc_weight_mcmc}
\end{equation}
This will be the situation throughout the remainder of this paper.

\subsection{Sequential Monte Carlo for Model Comparison}

The problem of interest is characterising the posterior distribution over
$\{\Mk\}_{k\in\mset}$, a set of possible models, with model $\Mk$ having
parameter vector $\paramk\in\Paramk$ which must also usually be inferred.
Given prior distributions $\pi(\Mk)$ and $\pi(\paramk|\Mk)$ and likelihood
$p(\data|\paramk,\Mk)$ we seek the posterior distributions $\pi(\Mk|\data)
\propto p(\data|\Mk)$. There are three fundamentally different approaches to
the computations:
\begin{enumerate}[\hspace*{1cm} \bf 1.]
  \item Calculate posterior model probabilities directly.
  \item Calculate the evidence, $p(\data|\Mk)$, of each model.
  \item Calculate pairwise evidence ratios.
\end{enumerate}
Each approach admits a natural \smc strategy. The relative strengths of these
approaches %, which are introduced in the following sections,
and alternative methods are identified in Table~\ref{tab:algadv}.

\begin{table}
\centering \footnotesize
\begin{tabularx}{0.8\linewidth}{Xccccccc}
  \toprule
  & \begin{sideways} \phm           \end{sideways}
  & \begin{sideways} \rjmcmc        \end{sideways}
  & \begin{sideways} \pmcmc         \end{sideways}
  & \begin{sideways} \textbf{\smcone} \end{sideways}
  & \begin{sideways} \textbf{\smctwo} \end{sideways}
  & \begin{sideways} \textbf{\smcthree} \end{sideways} \\
  \midrule
  \small Can deal with a countable set of models
  &       & \tick &       & \tick &       &       \\
  \small Can exploit inter-model relationships
  &       & \tick &       & \tick &       & \tick \\
  \small Characterises improbable models
  & \tick &       & \tick &       & \tick & \tick \\
  \small Doesn't require reversible-pairs of moves
  & \tick &       & \tick & \tick & \tick & \tick \\
  \small Doesn't require inter-model mixing
  & \tick &       & \tick &       & \tick &       \\
  \small Admits straightforward parallelisation
  &       &       & \tick/\cross  & \tick & \tick & \tick\\
  \small Doesn't rely upon ergodicity arguments
  &       &       &       & \tick & \tick & \tick \\
  \bottomrule
\end{tabularx}
\caption{Strengths of algorithms for model choice. \pmcmc admits
  a degree of parallelisation, but is not a natural candidate for
  implementation on massively-parallel architectures.}
\label{tab:algadv}
\end{table}

\subsubsection{\smcone: An All-in-One Approach}

One could consider obtaining samples from the same distribution employed in
the \rjmcmc approach to model comparison, namely:
\begin{equation}
  \pi^{(1)}(\Mk,\paramk) \propto \pi(\Mk)\pi(\paramk|\Mk)p(\data|\paramk,\Mk)
\end{equation}
which is defined on the disjoint union space
$\bigcup_{k\in\mset}(\{\Mk\}\times\Paramk)$.

One obvious \smc approach is to define a sequence of distributions
$\{\pi_t^{(1)}\}_{t=0}^T$ such that $\pi^{(1)}_0$ is easy to sample from,
$\pi_{T}^{(1)} = \pi^{(1)}$ and the intermediate distributions move smoothly
between them. In the remainder of this section, we use the notation
$(\Mk[t],\paramk[t])$ to denote a random sample on the space
$\bigcup_{k\in\mset}(\{\Mk\}\times\Paramk)$ at time $t$. One simple approach
is the use of an annealing scheme such that:
\begin{equation}
  \pi^{(1)}_t(\Mk[t],\paramk[t]) \propto \pi(\Mk[t])\pi(\paramk[t]|\Mk[t])
  p(\data|\paramk[t],\Mk[t])^{\alpha(t/T)},
  \label{eq:geometry_1}
\end{equation}
for some monotonically increasing $\alpha:[0,1]\to[0,1]$ such that $\alpha(0)
= 0$ and $\alpha(1) = 1$. Other approaches are possible and might prove more
efficient for some problems (such as the ``data tempering'' approach that
\citet{Chopin:2002hg} proposed for parameter estimation% which could easily
%be incorporated in our framework
--- a strategy which would lend itself
naturally to ``online'' estimation of evidence, but which would preclude the
use of the path sampling estimator), but this strategy provides a convenient
generic approach. These choices lead to Algorithm~\ref{alg:smc1}.

% $\alpha$ provides an \emph{annealing schedule} which is intended to allow
% the control of the discrepancy between pairs of adjacent distributions. This
% gives the joint prior on model order and parameters as an initial
% distribution and gradually introduces the influence of the likelihood until
% the full posterior is reached.

% Given a weighted collection of samples $\{W^{(k,i)}_T, X^{(k,i)}_{T} =
% (K_T^{(k,i)},\theta_T^{(k,i)})\}_{i=1}^N$ targeting $\pi^{(1)}$, one may
% approximate the two distributions of interest using the empirical measures:
% \begin{align*}
%   \widehat{\pi}^{(1)}(K = k) &= \sum\limits_{i=1}^N W^{(k,i)} \delta_{k,
%   K^{(k,i)}_T} \\
%   \widehat{\pi}^{(1)}(d\theta|K=k) &= \frac{1}{\widehat{\pi}^{(1)}(K = k)}
%   \sum\limits_{i=1}^N W^{(k,i)} \delta_{k, K_T^{(k,i)}}
%   \delta_{\theta^{(k,i)}_T}(d\theta),
% \end{align*}
% where $\delta_{k,K^{(k,i)}_T}$ denotes the \emph{Kronecker} delta and
% $\delta_{\theta{(k,i)}}$ denotes the Dirac measure concentrated at
% $\theta^{(k,i)}$.

% Thus, the output of such an \smc algorithm may be used as a (weighted)
% equivalent to that from a suitable \rjmcmc algorithm, with the posterior
% probability of each model corresponding to the (weighted) proportion of
% samples from that model.

This approach might outperform \rjmcmc when it is difficult to design
fast-mixing Markov kernels. Such an an \smc
strategy can outperform \mcmc at a given computational cost --- see, for
example, \citet{Fan:2008tf,Johansen:2008kp,Fearnhead:2010ua}. Such
trans-dimensional \smc has been proposed in several contexts
\citep{Peters:2005wh} and an extension proposed and analysed by
\citet{Jasra:2008bb}.

\begin{algorithm}
\small
\begin{algorithmic}
  \STATE \emph{Initialisation:} Set $t\leftarrow0$.
  \STATE\STATESKIP Sample $X_0^{(i)} = (M_0^{(i)},\theta_0^{(i)})\sim\nu$
  for some proposal distribution $\nu$ (usually the joint prior).
  \STATE\STATESKIP Weight $W_0^{(i)} \propto w_0(X_0^{(i)}) =
  {\pi(M_0^{(i)}) \pi(\theta^{(i)}_0|M_0^{(i)})}/
  {\nu(M_0^{(i)},\theta_0^{(i)})}$.
  \STATE\STATESKIP Apply resampling if necessary (e.g., if \ess
  \citep{Kong:1994ul} less than some threshold).

  \STATE \emph{Iteration:} Set $t\leftarrow t + 1$.
  \STATE\STATESKIP Weight $W_t^{(i)} \propto W_{t-1}^{(i)}
  p(\data|\theta_{t-1}^{(i)},M_{t-1}^{(i)})^{\alpha(t/T) - \alpha([t-1]/T)}$.
  \STATE\STATESKIP Apply resampling if necessary.
  \STATE\STATESKIP Sample $X_t^{(i)} \sim K_t(\cdot|X_{t-1}^{(i)})$, a
  $\pi_t^{(1)}$-invariant kernel.

  \STATE \emph{Repeat} the \emph{Iteration} step \emph{until $t = T$}.
\end{algorithmic}
\caption{\smcone: An All-in-One Approach to Model Comparison.}\label{alg:smc1}
\end{algorithm}

We include this approach for completeness and study it empirically later. Like
other trans-dimensional methods, this approach depends upon collection of
models being specified in advance. If new models are considered, then the
entire simulation must be redone. The more direct approaches described in the
following sections lead more naturally to easy-to-implement strategies with
good performance.

\subsubsection{\smctwo: A Direct-Evidence-Calculation Approach}
An alternative approach would be to estimate explicitly the evidence
associated with each model. We propose to do this by sampling from a sequence
of distributions for each model: starting from the parameter prior and
sweeping through a sequence of distributions to the posterior.

Numerous strategies are possible to construct such a sequence of
distributions, but one option is to use for each model $\Mk$, $k\in\mset$, the
sequence $\{\pi_t^{(2,k)}\}_{t=0}^{T_k}$, defined by
\begin{equation}
  \pi_t^{(2,k)}(\paramk[t]) \propto
  \pi(\paramk[t]|\Mk)p(\data|\paramk[t],\Mk)^{\alpha_k(t/T_k)}.
  \label{eq:geometry_2}
\end{equation}
where the number of distribution $T_k$, and the annealing schedule,
$\alpha_k:[0,1]\to[0,1]$ may be different for each model. This leads to
Algorithm~\ref{alg:smc2}.

The estimator of the posterior model probabilities depends upon the approach
taken to estimate the normalizing constant. Direct estimation of the evidence
can be performed using the output of this \smc algorithm and the standard
estimator \citep[Equation~14]{DelMoral:2006hc}, termed
\smctwo-\ds below:
\begin{equation}\label{eq:smc2-ds}
  \frac{1}{N} \sum_{i=1}^N \frac{\pi(\theta_0^{(k,i)}|\Mk)}{\nu(\theta_0^{(k,i)})} \times
  \prod_{t=2}^T \sum_{i=1}^N W_{t-1}^{(k,i)}
  p(\data|\theta_{t-1}^{(k,i)}\Mk)^{\alpha_k(t/T_k) - \alpha_k([t-1]/T_k)}
\end{equation}
where $W_{t-1}^{(k,i)}$ is the importance weight of sample $i$,
$\theta_{t-1}^{(k,i)}$, \emph{after} any resampling step of iteration $t-1$ for
model $\Mk$. This formula can be simplified by replacing $W_{t-1}^{(k,i)}$
with $1/N$ when resampling is conducted at every iteration (in which case it
is unbiased); otherwise a mathematically simpler representation less naturally
suited to computational use is provided by
\citet[Equation~15]{DelMoral:2006hc}. An alternative approach to computing the
evidence is also worthy of consideration. As has been suggested, and shown
empirically to
perform well previously \citep[see, for example]{Johansen:2006wm},
it is possible to use all of the samples from every generation of an \smc
sampler to approximate the path sampling estimator. % and hence to obtain an
%estimate of the ratio of normalizing constants.
Section~\ref{sub:Path Sampling via smctwo/smcthree} provides details.

The posterior distribution of the parameters conditional upon a particular
model can also be approximated using:
\begin{equation*}
  \widehat{\pi}_{T_k}^{(2,k)}(\diff\theta) =
  \sum\limits_{i=1}^{N} W_{T_k}^{(k,i)}
  \delta_{\theta^{(k,i)}_{T_k}}(\diff\theta).
\end{equation*}
%where $\delta_{\theta^{(k,i)}_{T_k}}$ is the Dirac measure.

% It is straightforward to use different numbers of samples, $N$, for each
% model when appropriate. in settings in which some models can be
% characterised substantially more easily than others.

This approach is appealing for several reasons. It is designed to
estimate directly the quantity of interest: the evidence. It provides as good a
characterisation of each model as is required: it is possible to obtain a good
estimate of the parameters of every model, even those for which the posterior
probability is small (although, of course, in certain circumstances the
automatic assignment of computational resources to the most promising models
may be desirable). Perhaps most significant is that this approach does not
require the design of proposal distributions or Markov kernels which move from
one model to another: each model is dealt with in isolation. Whilst this may
not be desirable in every situation, there are circumstances in which
efficient moves between models are almost impossible to devise.

This approach also has some disadvantages. In particular, it is necessary to
run a separate simulation for each model --- rendering it impossible to deal
with countable collections of models (although this is not such a substantial
problem in many interesting cases). The ease of implementation may often
offset this limitation.

\begin{algorithm}
\small
\begin{algorithmic}
  \STATE For each model $k \in \mset$ execute the following algorithm.

  \STATE \emph{Initialisation:} Set $t\leftarrow0$.
  \STATE\STATESKIP Sample $\theta_0^{(k,i)}\sim\nu_k$ for some proposal
  distribution $\nu_k$ (usually the parameter prior).
  \STATE\STATESKIP Weight $W_0^{(k,i)} \propto w_0(\theta_0^{(k,i)}) =
  {\pi(\theta_0^{(k,i)}|\Mk)}/{\nu_k(\theta_0^{(k,i)})}$.
  \STATE\STATESKIP Apply resampling if necessary.

  \STATE \emph{Iteration:} Set $t\leftarrow t + 1$.
  \STATE\STATESKIP Weight $W_t^{(k,i)} \propto W_{t-1}^{(k,i)}
  p(\data|\theta_{t-1}^{(k,i)},\Mk)^{\alpha(t/T_k) - \alpha([t-1]/T_k)}$.
  \STATE\STATESKIP Apply resampling if necessary.
  \STATE\STATESKIP Sample $\theta_t^{(k,i)} \sim
  K_t(\cdot|\theta_{t-1}^{(k,i)})$, a $\pi_t^{(k,2)}$-invariant kernel.

  \STATE \emph{Repeat} the \emph{Iteration} step \emph{until $t = T_k$}.
\end{algorithmic}
\caption{\smctwo: A Direct-Evidence-Calculation Approach.}\label{alg:smc2}
\end{algorithm}

\subsubsection{\smcthree: A Relative-Evidence-Calculation Approach}

A final approach can be thought of as \emph{sequential model comparison}.
Rather than estimating the evidence associated with any particular model, we
could estimate pairwise evidence ratios directly. The \smc sampler starts with
an initial distribution being the posterior of one model (an initial sample
could be obtained using a secondary \smc algorithm or other sampler) and moves towards the
posterior of another related model. Then the sampler can continue towards
another related model and so forth.

Given a finite collection of models $\{\Mk\}$, $k\in\mset$, suppose the models
are ordered in a sensible way (e.g., $\Mk[k-1]$ is nested within $\Mk$ or
$\paramk$ is of higher dimension than $\paramk[k-1]$). For each
$k\in\mset$, we consider a sequence of distributions
$\{\pi_t^{(3,k)}\}_{t=0}^{T_k}$, such that $\pi_0^{(3,k)}(\Mk[],\paramk[]) =
\pi(\paramk[]|\data,\Mk) \mathbb{I}_{\{\Mk\}}(\Mk[])$ and $\pi_{T_k}^{(3,k)}(\Mk[],\paramk[]) =
\pi(\paramk[]|\data,\Mk[k+1]) \mathbb{I}_{\{\Mk[k+1]\}}(\Mk[]) = \pi_{0}^{(3,k+1)}(\Mk[],\paramk[])$.
When it is possible to construct a \smc sampler that iterates over this
sequence of distributions, the estimate of the ratio of normalizing constants
is the Bayes factor estimate of model $\Mk[k+1]$ in favour of model $\Mk$.

This approach is conceptually appealing, but requires the construction of a
smooth path between the posterior distributions of interest. The geometric
annealing strategy which has been advocated as a good generic strategy in the
previous sections is only appropriate when the support of successive
distributions is non-increasing. This is unlikely to be the case in
interesting model comparison problems.

In this paper we consider a sequence of distributions on the disjoint union
$\{\Mk,\Paramk\}\cup\{\Mk[k+1],\Paramk[k+1]\},$ with the sequence of
distributions $\{\pi_t^{(3,k)}\}_{t=0}^{T_k}$ defined as the full posterior,
\begin{equation}
  \pi_t^{(3,k)}(\Mk[t],\paramk[t]) \propto
  \pi_t(\Mk[t]) \pi(\paramk[t]|\Mk[t]) p(\data|\paramk[t],\Mk[t])
\end{equation}
where $\Mk[t]\in\{\Mk,\Mk[k+1]\}$ and the ``prior'' over models at time $t$,
$\pi_t(\Mk[k+1]) := \alpha(t/T_k)$,  for some monotonically increasing bijection
$\alpha:[0,1]\to[0,1]$. The \mcmc moves between need to be similar to those in the
\rjmcmc or \smcone algorithms. However, instead of efficient
exploration of the whole model space, only moves between two models are
required and the sequence of distributions employed helps to ensure
exploration of both model spaces. Algorithm~\ref{alg:smc3} uses this particular sequence
of distribution  but other sequence of distributions between models could be employed.

An advantage of this approach is that it provides direct estimates of the
Bayes factor which is of interest for model comparison purpose while not
requiring exploration of as complicated a space as that employed within
\rjmcmc or \smcone. The estimation of normalizing constant in \smcthree
follows in exactly the same manner as in the \smctwo case. In \smcthree, the
same estimator provides a direct estimate of the Bayes factor.

\begin{algorithm}
\small
\begin{algorithmic}
  \STATE \emph{Initialisation:} Set $k\leftarrow1$.
  \STATE\STATESKIP Use Algorithm~\ref{alg:smc2} to obtain weighted samples
  for $\pi_{T_1}^{(3,1)}$, the parameter posterior for model $\Mk[1]$

  \STATE \emph{Relative Evidence Calculation}
  \STATE\STATESKIP Set $k\leftarrow k + 1$, $t\leftarrow0$.
  \STATE\STATESKIP Denote current weighted samples as
  $\{W_0^{(k,i)},X_0^{(k,i)}\}_{i=1}^N$ where $X_0^{(k,i)} =
  (M_0^{(k,i)},\theta_0^{(k,i)})$
  \STATE\STATESKIP Apply resampling if necessary.

  \STATE\STATESKIP \emph{Iteration:} Set $t\leftarrow t + 1$.
  \STATE\STATESKIP\STATESKIP Weight $W_t^{(k,i)} \propto W_{t-1}^{(k,i)}
  {\pi_t(M_{t-1}^{(k,i)})}/{\pi_{t-1}(M_{t-1}^{(k,i)})}$.
  \STATE\STATESKIP\STATESKIP Apply resampling if necessary.
  \STATE\STATESKIP\STATESKIP Sample $(M_t^{(k,i)},\theta_t^{(ki)}) \sim
  K_t(\cdot|M_{t-1}^{(k,i)}\theta_{t-1}^{(k,i)})$, a $\pi_t^{(3,k)}$-invariant
  kernel.

  \STATE\STATESKIP \emph{Repeat the \emph{Iteration} step up to $t = T_k$}.

  \STATE \emph{Repeat} the \emph{Relative Evidence Calculation} step \emph{until
    sequentially all relative evidences are calculated}.
\end{algorithmic}
\caption{\smcthree: A Relative-Evidence-Calculation Approach to Model
  Comparison.}
\label{alg:smc3}
\end{algorithm}

\subsection{Path Sampling via \smctwo/\smcthree}
\label{sub:Path Sampling via smctwo/smcthree}

%The estimation of the normalizing constant associated with our sequences of
%distributions can be achieved by a
Monte Carlo approximation to the \emph{path
  sampling} identity \citep{Gelman:1998ei} (also known as thermodynamic
integration or Ogata's method) also provides an estimate of the normalising constant. The
use of \ais for the same purpose \citep{Neal:2001we} is common in some
settings; as will be demonstrated below the incorporation of some other
elements of the more general \smc algorithm family can improve performance at
negligible cost. Given a parameter $\alpha$ which defines a family of distributions,
$\{p_{\alpha} = q_{\alpha} / Z_\alpha\}_{\alpha \in [0,1]}$ which move
smoothly from $p_0 = q_0 / Z_0$ to $p_1 = q_1 / Z_1$ as $\alpha$ increases
from zero to one. The logarithm of the ratio of their normalizing constants satisfies a simple integral relationship
under mild regularity conditions:
\begin{equation}
  \log\biggl( \frac{Z_1}{Z_0} \biggr) =
  \int_{0}^{1} \Exp_\alpha \biggl[ \rnd{\log q_{\alpha}(\cdot)}{\alpha}
  \biggr] \intd\alpha, \label{eq:path_identity}
\end{equation}
where $\Exp_\alpha$ denotes expectation under $p_\alpha$; see
\citet{Gelman:1998ei}. Note that the sequence of distributions in the \smctwo
and \smcthree algorithms above, can both be interpreted as belonging to such a
family of distributions, with $\alpha_t = \alpha(t/T_k)$, where the mapping
$\alpha:[0,1]\to[0,1]$ is again monotonic with $\alpha(0) = 0$ and $\alpha(1)
= 1$.

The \smc sampler provides us with a set of weighted samples obtained from a
sequence of distributions suitable for approximating this integral. At each
$t$ we can obtain an estimate of the expectation within the integral for
$\alpha(t/T)$ via the usual importance sampling estimator, and this integral
can then be approximated via numerical integration. Whenever the sequence
of distributions employed by \smcthree has appropriate differentiability it is
also possible to employ path sampling to estimate, directly, the evidence
ratio via this approach applied to the samples generated by that algorithm. In
general, given an increasing sequence $\{\alpha_t\}_{t=0}^T$ where $\alpha_0 =
0$ and $\alpha_T = 1$, a family of distributions
$\{p_{\alpha}\}_{\alpha\in[0,1]}$ as before, and a \smc sampler that iterates
over the sequence of distribution $\{\pi_t = p_{\alpha_t} =
q_{\alpha_t}/Z_{\alpha_t}\}_{t=0}^T$, then with the weighted samples
$\{W_t^{(j)},X_t^{(j)}\}_{j=1}^N$, and $t = 0,\dots,T$, a path sampling
estimator of the ratio of normalizing constants $\Xi_T = \log(Z_1/Z_0)$
can be approximated (using an elementary trapezoidal scheme) by
\begin{equation}
  \widehat\Xi_{T}^{N} = \sum_{t=1}^T
  \frac{1}{2}(\alpha_t - \alpha_{t - 1})(U_t^N + U_{t-1}^N)
  \label{eq:path_est}
\end{equation}
where
\begin{equation}
  U_t^N = \sum_{j=1}^N
  W_t^{(j)} \rnd{\log q_{\alpha}(X_t^{(j)})}{\alpha}\Bigm|_{\alpha = \alpha_t}.
\end{equation}

We term these estimators \smctwo-\ps and \smcthree-\ps. The
combination of \smc and path sampling is somewhat natural and has been
proposed before, e.g., \citet{Johansen:2006wm} although not there in a Bayesian
context. %Despite the good performance observed in the setting of rare
%event simulation, t
The estimation of normalizing constants by this approach seems to have
received little attention in the literature. Perhaps because of widespread
acceptance of the suggestion of \citet{DelMoral:2006hc}, that \smc doesn't
outperform \ais when normalizing constants are the object of inference or that
of \citet{Calderhead:2009bd} that all simulation-based estimators based around
path sampling can be expected to behave similarly. We will demonstrate below that these
observations, whilst true in certain contexts, do not hold in full generality.

\subsection{Extensions and Refinements}
\label{sub:Extensions and Refinements}

\subsubsection{Improved Univariate Numerical Integration}
\label{ssub:Improved Univariate Numerical Integration}
The path sampling estimator requires evaluation of
the expectation,
$  \Exp_\alpha [ {\diff \log q_{\alpha}}/ {\diff \alpha}]$
for $\alpha\in[0,1]$, which can be approximated by importance sampling using
samples generated by a \smc sampler operating on the sequence of distributions
$\{\pi_t = p_{\alpha_t} = q_{\alpha_t}/Z_t\}_{t=0}^T$ directly for
$\alpha\in\{\alpha_t\}_{t=0}^T$. For any $\alpha\in[0,1]$, by finding $t$
such that $\alpha\in(\alpha_{t-1},\alpha_t)$, the expectation can be %easily
approximated using existing \smc samples --- the quantities required to obtain
such an estimate have already been calculated during the running of the \smc
algorithm and such computations have little computational cost.
%Indeed, in settings in which the \smc importance weights
%for target distribution $\pi_{\alpha}$ depend only upon the sample at time
%$t-1$, as in the proposed algorithms, the computational cost of the new
%weights are often minimal compared to generate new samples.

As noted by \cite{Friel:2012}  we can use more sophisticated numerical
integration strategies to reduce the path sampling estimator bias. %For
%example, higher order Newton-Cotes rules rather than the Trapezoidal rule can
%be implemented straightforwardly.
In the case of SMC it is especially straightforward to estimate the required
expectations at arbitrary $\alpha$ and so %we cheaply use
higher order integration can be used cheaply. Numerical integrations which make use of a finer
mesh $\{\alpha_t'\}_{t=0}^{T'}$ than $\{\alpha_t\}_{t=0}^T$ can be easily implemented.
Due to the possibile instability of numerical integrations based on
approximations of derivatives,
%obtained from Monte Carlo methods may potentially be unstable in some
%situations,
the second approach can be more appealing in some applications. A demonstration of the
bias reduction effect is provided in Section~\ref{sub:Positron Emission
  Tomography Compartmental Model}.

\subsubsection{Adaptive Specification of Distributions}
\label{ssub:Adaptive Specification of Distributions}

As the importance weights at time $t$ depend only upon the sample at time
$t-1$, it is relatively straightforward to consider sample-dependent, adaptive
specification of the sequence of distributions (typically by choosing the
value of a parameter, such as $\alpha_t = \alpha(t/T_k)$ in the
settings of \smctwo and \smcthree, based upon the current sample).
\citet{Jasra:2010eh} proposed such a method based on controlling the rate at
which the
effective sample size (\ess; \citet{Kong:1994ul}) falls. With little
computation cost, this provides an automatic method of specifying a tempering
schedule in such a way that the \ess decays in a regular fashion.
\citet[Algorithm 2]{Schafer:2011bx} used a similar technique but by moving the
particle system only when it resamples they are in a setting
equivalent to resampling at every timestep (with longer time steps, followed
by multiple applications of the MCMC kernel) in our formulation. We advocate
resampling adaptively only when the \ess is smaller than a preset threshold,
and here we propose a more general adaptive scheme for the selection of the
sequence of distributions which has better properties when
adaptive resampling is employed.

The \ess was designed to assess the loss of efficiency arising from the use a
simple weighted sample (rather than a simple random sample from the
distribution of interest) in the computation of expectations. It is obtained
by considering a sample approximation of a low order Taylor expansion of the
variance of the importance sampling estimator of an arbitrary test function to
that of the simple Monte Carlo estimator; the test function vanishes from the
expression as a consequence of this expansion.

In our context, allowing $W_{t-1}^{(i)}$ to denote the \emph{normalized
  weights} of particle $i$ at the end of time $t - 1$, and $w_t^{(i)}$ to
denote the \emph{unnormalized} incremental weights of particle $i$ during
iteration $t$ the \ess calculated using the current weight of each particle is
simply:
\begin{equation}
  \ess_t = \left[ {\sum_{j=1}^N\left( \frac{W_{t-1}^{(j)}
          w_t^{(j)}}{\sum_{k=1}^NW_{t-1}^{(k)}w_t^{(k)}}\right)^2}
  \right]^{-1} = \frac{\bigl(\sum_{j=1}^NW_{t-1}^{(j)}w_t^{(j)}\bigr)^2}
  {\sum_{k=1}^N\bigl(W_{t-1}^{(k)}\bigr)^2\bigl(w_t^{(k)}\bigr)^2}.
\end{equation}
It is clearly appropriate to use this quantity (which corresponds to the
coefficient of variation of the current normalized importance weights) to
assess weight degeneracy and to make decisions about appropriate resampling
times (cf. \cite{DelMoral:2012jq}) but it is rather less apparent that it is
the correct quantity to consider when adaptively specifying a sequence of
distributions in an \smc sampler.

The \ess of the current sample weights tells us about the accumulated mismatch
between proposal and target distributions (on an extended space including the
full trajectory of the sample paths) since the last resampling time. Fixing
either the relative or absolute reduction in \ess between successive
distributions does \emph{not} lead to a common discrepancy between successive
distributions unless resampling is conducted after every iteration as will be
demonstrated below.

When specifying a sequence of distributions it is natural to aim for a similar
discrepancy between each pair of successive distributions. %In the context of
%effective sample size,
The natural question to ask is consequently, how large can we make $\alpha_t - \alpha_{t-1}$ whilst ensuring that $\pi_{t}$ remains
sufficiently similar to $\pi_{t-1}$. One way to measure the discrepancy would
be to consider how good an importance sampling proposal $\pi_{t-1}$ would be
for the estimation of expectations under $\pi_t$ and a natural way to measure
this is via the sample approximation of a Taylor expansion of the relative
variance of such an estimator exactly as in the \ess.

Such a procedure (see the supplementary material for its derivation) leads us
to a quantity which we have termed the \emph{conditional} \ess (\cess):
\begin{equation}
  \cess_t = \left[ {\sum_{j=1}^N N W_{t-1}^{(j)} \left(
        \frac{w_t^{(j)}}{\sum_{k=1}^N
          N W_{t-1}^{(k)}w_t^{(k)}}\right)^2} \right]^{-1}
  = \frac{N \bigl(\sum_{j=1}^NW_{t-1}^{(j)}w_t^{(j)}\bigr)^2}
  {\sum_{k=1}^N W_{t-1}^{(k)} \bigl(w_t^{(k)}\bigr)^2}
\end{equation}
%% \begin{equation}
%%   \cess_n =
%% \frac
%%   {\bigl(\sum_{j=1}^N W_{t-1}^{(j)}w_t^{(j)}\bigr)^2}
%%   {\sum_{j=1}^N W_{t-1}^{(j)}\bigl(w_t^{(j)}\bigr)^2}
%% =
%% \frac
%%   {\bigl(\sum_{j=1}^NW_{t-1}^{(j)}w_t^{(j)}\bigr)^2}
%%   {\sum_{j=1}^N \frac{1}{N} W_{t-1}^{(j)}\bigl(w_t^{(j)}\bigr)^2}
%% \end{equation}
which is equal to the \ess only when resampling is conducted during every
iteration. The bracketed term coincides with a sample approximation (using the
actual sample which is properly weighted to target $\pi_{t-1}$) of the
expected sum of the unnormalized weights squared divided by the square of a
sample approximation of the expected sum of unnormalized weights when
considering sampling from $\pi_{t-1}$ and targeting $\pi_t$ by simple
importance sampling.

\begin{figure}
  \center\includegraphics[width=0.85\textwidth]{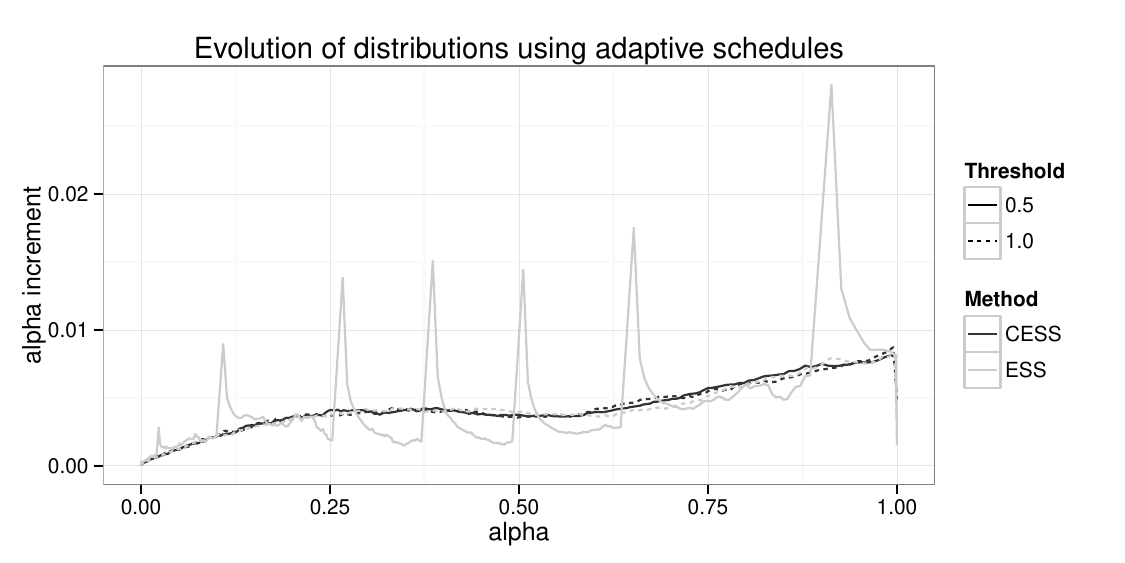}
  \caption{A typical plot of $\alpha_t - \alpha_{t-1}$ against $\alpha_t$ (for
    a Gaussian mixture model example using the \smctwo algorithm; see the
    supplementary material). %The specifications of the adaptive
    %parameter (\ess or \cess) are adjusted such that
    All four samplers use roughly the same number of distributions.}
  \label{fig:adaptive_alpha}
\end{figure}

Figure~\ref{fig:adaptive_alpha} shows the variation of $\alpha_t -
\alpha_{t-1}$ with $\alpha_t$ when fixed reductions in \ess and \cess are used
to specify the sequence of distributions both when resampling is conducted
during every iteration (or equivalently, when the $\ess/N$ falls below a
threshold of 1.0) and when resampling is conducted only when the $\ess/N$
falls below a threshold of 0.5. As is demonstrated in
Section~\ref{sec:Illustrative Applications} the \cess-based scheme leads to a
reduction in estimator variance of around 20\% relative to a manually tuned
(quadratic; see the supplementary material) schedule while the \ess-based
strategy provides little improvement over the linear case unless resampling is
conducted during every iteration.

In addition to providing a significantly better performance at essentially no
cost, the use of the \cess emphasizes the purpose of the adaptive
specification of the sequence of distributions: to produce a sequence in which
the difference between each successive pair is the same (when using the \cess
one is seeking to ensure that the variance of the importance weights one would
arrive at if using $\pi_{t-1}$ as a proposal for $\pi_t$ is constant).

We note that the standard estimate of the normalising constant need not be
unbiased when adaptive techniques are employed. However, a very recent
analysis \citep{smc:clt:BJT13} provides some formal justification of the use
of both adaptive tempering schedules and adaptive specification of proposals,
the topic of the next section.

\subsubsection{Adaptive Specification of Proposals}
\label{ssub:Adaptive Specification of Proposals}

The \smc sampler is remarkably robust to the mixing speed of \mcmc kernels
employed (see the empirical study below). However, as with any
sampling algorithms, faster mixing doesn't harm performance and in some cases
will considerably improve it. For random walk Metropolis kernels, the mixing
speed depends upon the proposal \emph{scale}.

We adopt a similar approach to \citet{Jasra:2010eh} who
%. They applied an
%idea used within adaptive \mcmc methods \citep{Andrieu:2006tw} to \smc
%samplers by using variance of parameters estimated from its particle system
use sample covariance estimates to inform the proposal covariance for the next iteration.
%Although, in practice
We found that such an approach generally produces satisfactory results and it
is simple to implement. In difficult problems alternative approaches could be employed; one approach demonstrated in
\citet{Jasra:2010eh} is to simply employ a pair of acceptance rate thresholds
and to alter the proposal scale from the simply estimated value whenever the
acceptance rate falls outside those threshold values. In
\citet{smc:clt:BJT13}, convergence results were shown for this kind of
adaptive specification of Markov kernels.

More sophisticated proposal strategies could undoubtedly improve performance
further and their use warrants investigation. One appealing approach is using
the Metropolis adjusted Langevin algorithm (\mala; see
\citet{Roberts:1996vd}).
%  In summary, \mala derives a Metropolis-Hastings
% proposal kernel for a target $\pi$ which satisfies suitable differentiability
% and positivity conditions, from the Langevin diffusion,
% \begin{equation*}
%   \diff L_t = \frac{1}{2}\nabla\log\pi(L_t)\diff t + \diff B_t
% \end{equation*}
% where $B_t$ is the standard Brownian motion. Given a state $X_{n-1}$, a new
% state is proposed by discrete approximation to the above diffusion. That is,
% for a fixed $h>0$,
% \begin{equation}
%   X_n\sim\rnorm\Bigl(X_{n-1}+\frac{1}{2}\nabla\log\pi(X_{n-1}), hI_d\Bigr)
% \end{equation}
% where $I_d$ is the identify matrix and $d$ is the dimension of the state
% space. The new proposed state is accepted or rejected through the usual
% Metropolis-Hastings algorithm.
%Compared to a ``vanilla'' random walk \mala is attractive.% when it is
%possible and its convergence conditions \citep{Roberts:1996vd} can be met,
%because only one discrete approximation parameter $h$ needs to be tuned
%for optimal performance. In addition, results from \citet{Roberts:2001ta}
%suggested that \mala can be more efficient than a random walk when using
%optimal scalings.
We could use the particle approximation at time index
$t = n - 1$ to estimate the covariance matrix of $\pi_{n}$ and thus tune the
scale $h$ on-line. As these algorithms are known to be somewhat sensitive to
scaling, and we seek approaches robust enough to employ with little user
intervention, we have not investigated this strategy here.

\subsection{A Near-Automatic, Generic Algorithm}
\label{sub:An Automatic, Generic Algorithm}

With the above refinements, the \smctwo algorithm can be implemented with
minimal tuning and application-specific effort while providing robust and
accurate estimates of the model evidence $p(\data|\Mk)$. The geometric
annealing path that connects the prior $\pi(\paramk|\Mk)$ and the posterior
$\pi(\paramk|\data,\Mk)$, provides a smooth path for a wide range of problems.
The actual annealing schedule under this scheme can be determined
using the adaptive schedule as described above.  Finally, we can adaptively
specify the Metropolis random walk (or \mala) scales through the estimation of
their scaling parameters as the sampler iterates. In contrast to the \mcmc
setting, where such adaptive algorithms will usually require
a burn-in period, which will not be used for further estimation, in \smc, the
variance and covariance estimates come at almost no cost, as all the samples
will later be used for marginal likelihood estimation. Additionally,
adaptation within \smc does not require separate theoretical justification ---
something which can significantly complicate the development of adaptive
schemes in the \mcmc setting. %Alternatively, we can
%also specify the proposal scales in a deterministic, but sensible way. Since
%\smc algorithms are relatively robust to the change of scales, such
%deterministic scales will not require the same degree of tuning as is required
%to obtain good performance in \mcmc algorithms.
%The adaptive strategy can be applied to both algorithms directly. The
%applicability to the \smcthree algorithm depends on the nature of the sequence
%of distributions.
We outline the adaptive form of \smctwo in Algorithm~\ref{alg:adaptive}.

\begin{algorithm}
\small
\begin{algorithmic}
  \STATE \emph{Accuracy control}
  \STATE\STATESKIP Set constant $\cess^\star\in(0,1)$, using a small pilot
  simulation if necessary.

  \STATE \emph{Initialization:} Set $t\leftarrow0$.
  \STATE\STATESKIP Perform the \emph{Initialization} step
    as in Algorithm~\ref{alg:smc2}

  \STATE \emph{Iteration:} Set $t\leftarrow t + 1$

  \STATE\STATESKIP \emph{Step size selection}
  \STATE\STATESKIP\STATESKIP Use a binary search %(or other search algorithms)
  to find $\alpha^\star$ such that $\cess_{\alpha^\star} = \cess^\star$
  \STATE\STATESKIP\STATESKIP Set $\alpha_t \leftarrow\alpha^\star$ if
    $\alpha^\star \le 1$, otherwise set $\alpha_t\leftarrow1$

  \STATE\STATESKIP \emph{Proposal scale calibration}
  \STATE\STATESKIP\STATESKIP
  Computing the importance sampling estimates of first two moments of
  parameters.
  \STATE\STATESKIP\STATESKIP
  Set the proposal scale of the Markov proposal $K_t$ with the estimated
  parameter variances.

  \STATE\STATESKIP Perform the \emph{Iteration} step as in
  Algorithm~\ref{alg:smc2} with the found $\alpha_t$
  and proposal scales.

  \STATE \emph{Repeat} the \emph{Iteration} step %up to $t = T$ (or $T_k$ in the
    %case of Algorithm~\ref{alg:smc2})}
    \emph{until $\alpha_t = 1$} then set $T=t$.
\end{algorithmic}
\caption{An Automatic, Generic Algorithm for Bayesian Model Comparison}
\label{alg:adaptive}
\end{algorithm}

As laid out above, the algorithm requires minimal tuning. Its robustness,
accuracy and efficiency will be shown empirically in
Section~\ref{sec:Illustrative Applications}. Automating \smcone is less
straightforward as the between model moves still require effort to design and
implement. In \smcthree, the specification of the sequences between posterior
distributions are less generic than the geometric annealing scheme in
\smctwo. However, the adaptive schedule and automatic tuning of \mcmc proposal
scales can readily be applied.

Some auxiliary inputs are still required. However,
for a given class of models, with minimal tuning, the algorithm can be carried
out in a nearly automatic fashion for different data or model settings, in the
sense that these inputs do not need to be done on a per model or per data set
basis. We believe this framework presented here is at least a good foundation for
building automatic model comparison procedures for many application areas.

Although further enhancements and refinements are clearly possible, we focus
in the remainder of this article on this simple, generic algorithm which can
be easily implemented in any application and has proved sufficiently powerful
to provide good estimation in the examples we have encountered thus far.

\section{Illustrative Applications}
\label{sec:Illustrative Applications}

A classical Gaussian mixture model (\gmm) as formulated in
\citet{DelMoral:2006hc} was first used to compare all three \smc algorithms
with \rjmcmc, \ais and \pmcmc. The details of model setting and results are in
the supplementary material. It was found that all five algorithms agree on the
results while the performance in terms of Monte Carlo variance varies
considerably. We reached the conclusion that the \smctwo algorithm with
adaptive strategies is the most promising among the \smc strategies,
considering ease of implementation, performance and generality. Also, while it has been suggested that \ais might perform similarly to \smc for the estimation of normalising constants, the \gmm example shows that resampling can have a beneficial effect on the variance allowing \smc to outperform \ais in practice.

In this section, two realistic examples, a nonlinear \ode model and a Positron
Emission Tomography compartmental model are used to study the performance and
robustness of algorithm \smctwo compared to \ais and \pmcmc. Various
configurations of the algorithms are considered including both sequential and
parallelized implementations.

The \texttt{C++} implementations, which make use of the vSMC library of
\cite{software:VSMC}, of all examples can be found at
\url{https://github.com/zhouyan/vSMC}.

\subsection{Nonlinear Ordinary Differential Equations}

In this section, \smctwo will now be further
explored in a more complex model, a nonlinear ordinary differential equations
system. This model, which was studied in \cite{Calderhead:2009bd}, is known as
the Goodwin model. The \ode system, for an $m$-component model, is:
\begin{align*}
  \frac{\diff X_1(t)}{\diff t} &= \frac{a_1}{1 + a_2 X_m(t)^{\rho}}
  - \alpha X_1(t)  & \\
  \frac{\diff X_i(t)}{\diff t} &= k_{i-1}X_{i-1}(t) - \alpha X_i(t)
  & i = 2,\dots,m \\
  X_i(0) &= 0 & i = 1,\dots,m
\end{align*}
The parameters $\{\alpha,a_1,a_2,k_{1:m-1}\}$ have common prior distribution
$\rgamma(0.1, 0.1)$. Under this setting, $X_{1:m}(t)$ can exhibit either
unstable oscillation or a constant steady state. The data are simulated for
$m=\{3,5\}$ at equally spaced time points from $0$ to $60$, with time step
$0.5$. The last $80$ data points of $(X_1(t), X_2(t))$ are used for inference.
Normally-distributed noise with standard deviation $\sigma=0.2$ is added to
the simulated data. Following \cite{Calderhead:2009bd}, the variance of the
additive measurement error is assumed to be known. Therefore, the posterior
distribution has $m+2$ parameters for an $m$-component model.

As shown in \cite{Calderhead:2009bd}, when $\rho > 8$, due to the possible
instability of the \ode system, the posterior can have a considerable number
of local modes. In this example, we set $\rho = 10$. Also, as the solution to
the \ode system is somewhat unstable, slightly different data can result in
very different posterior distributions.

\subsubsection{Results}

We compare results from the \smctwo and \pmcmc algorithms. For the \smc
implementation, $1,000$ particles and $500$ iterations were used, with the
distributions specified by Equation~\eqref{eq:geometry_2}, with $\alpha(t/T) =
(t/T)^5$, or via the completely adaptive specification. For the \pmcmc
algorithm, $50,000$ iterations are performed for burn-in and another $10,000$
iterations are used for inference. The same tempering as was used for \smc is
used here. Note that, in a sequential implementation of \pmcmc, with each
iteration updating one local chain and attempting a global exchange, the
computational cost of after burn-in iterations is roughly the same as the
entire \smc algorithm. In addition, changing $T$ within the range of the
number of cores available does not substantially change the computational cost
of a generic parallel implementation of the \pmcmc algorithm, with each
iteration updating all local chains concurrently. We compare results from $T =
10,30,100$ for \pmcmc and $T = 500$ (or close to this number when the
distributions are specified adaptively) for \smc. The results for
data generated from the simple model ($m = 3$) and complex model ($m = 5$),
summarising variability amongst 100 runs of each algorithm, are shown in
Tables~\ref{tab:node-s-all} and~\ref{tab:node-c-all}, respectively.

\input{tab/node-s-all-ja}
\input{tab/node-c-all-ja}

As shown in both cases, the number of distributions can affect the performance
of \pmcmc algorithms considerably. When using $10$ distributions, large bias
from numerical integration for path sampling estimator was observed, as
expected. With $30$ distributions, the performance is comparable to the
\smctwo sampler, though some bias is still observable. With $100$
distributions, there is a much larger variance because, with more chains, the
information travels more slowly from rapidly mixing chains to slowly mixing
ones and consequently the mixing of the overall system is inhibited.

The \smc algorithm provides results comparable to the best of three \pmcmc
implementations in all settings, including one in which both the annealing
schedule and proposal scaling were fully automatic, and significantly better
for the data generated from simple model. In fact, the completely adaptive
strategy was the most successful.

It can be seen that in contrast to the \pmcmc algorithm, the \smc algorithm
can increase the number of the distributions to reduce the bias of the
numerical integration for the path sampling estimator without increasing the
Monte Carlo variance.

% It can be seen that increasing the number of distributions not only reduces
% the bias of numerical integration for path sampling estimator, but also
% reduces the variance considerably. On the other hand increasing the number of
% particles can only reduce the variance of the estimates, in accordance with
% the central limit theorem (see \citet{DelMoral:2006hc} for the standard
% estimator and extensions for path sampling estimator, Proposition~1{} in the
% supplementary material) .

%Therefore, though there exists the trade-off between the number of particles
%and the number distributions, increasing either of them can always benefit
%the accuracy of the estimators.

\subsection{Positron Emission Tomography Compartmental Model}
\label{sub:Positron Emission Tomography Compartmental Model}

It is now interesting to compare the proposed algorithm with other
state-of-art algorithms using a realistic example.

Positron Emission Tomography (\pet) is a technique used for studying the brain
\emph{in vivo}, most typically when investigating metabolism or neuro-chemical
concentrations in either normal or patient groups. Given the nature and number
of observations typically recorded in time, \pet data is usually modeled with
linear differential equation systems. For an overview of \pet compartmental
models see \citet{Gunn:2002tf}. Given data $(y_1,\dots,y_n)^{\textrm{T}}$, an
$m$-compartmental model has generative form:
\begin{gather}
  y_j = C_T(t_j;\phi_{1:m},\theta_{1:m}) + \sqrt{
    \frac{C_T(t_j;\phi_{1:m},\theta_{1:m})}{t_j-t_{j-1}}}
  \varepsilon_j \\
  C_T(t_j;\phi_{1:m},\theta_{1:m}) = \sum_{i=1}^m
  \phi_i\int_0^{t_j}C_P(s)e^{-\theta_i(t_j-s)}\intd s
\end{gather}
where $t_j$ is the measurement time of $y_j$, $\varepsilon_j$ is additive
measurement error and input function $C_P$ is (treated as) known. The
parameters $\phi_1,\theta_1,\dots,\phi_m,\theta_m$ characterize the model
dynamics. See \citet{Zhou2013} for applications of Bayesian model comparison
for this class of models and details of the specification of the measurement
error. In the simulation results below, $\varepsilon_j$ are independently and
identically distributed according to a zero mean Normal distribution of
unknown variance, $\sigma^2$, which was included in the vector of model
parameters.

\begin{figure}
  \center\includegraphics[width=0.85\linewidth]{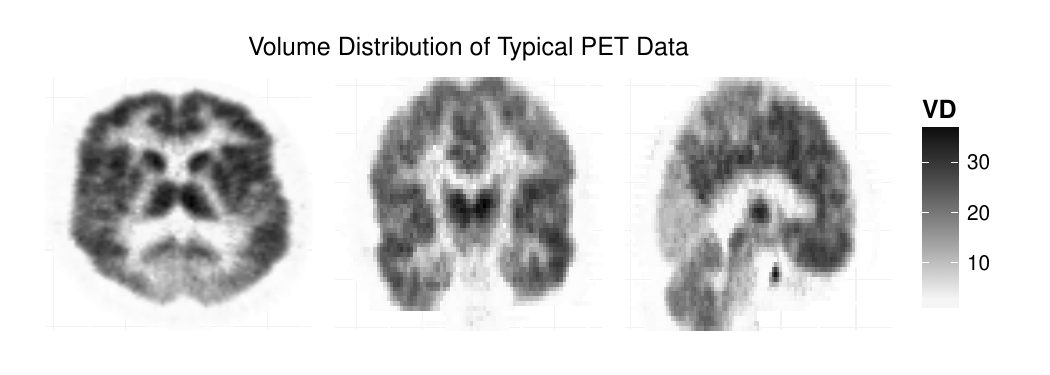}
  \caption{Estimates of $V_D$ from a single \pet scan as found using \smctwo.
    The data shows that the volume of distribution exhibits substantial
    spatial variation. Note that each pixel in the image represent an estimate
    from an individual time series. There are approximately 250,000
    of them and each requires a Monte Carlo simulation to select a
    model.}
  \label{fig:petplot}
\end{figure}

Real neuroscience data sets involve a very large number of time series ($\sim200,000$ per
brain), which are typically somewhat heterogeneous.
Figure~\ref{fig:petplot} shows estimates of $V_D =
\sum_{j=1}^m\phi_j/\theta_j$ from a typical \pet scan (generated using \smctwo
as will be discussed later). Robustness is therefore especially important. An
application-specific \mcmc algorithm was developed for this problem in
\citet{Zhou2013}. A significant amount of tuning of the algorithms was
required to obtain good results. The results shown in Figure~\ref{fig:petplot}
are very close to those of \citet{Zhou2013} but, as is shown later, they were
obtained with almost no manual tuning effort and at similar computational
cost.

For \smc and \pmcmc algorithms, the requirement of robustness means that the
algorithm must be able to calibrate itself automatically to different data
(and thus different posterior surfaces). A sequence of distributions which
performs well for one time series may not perform even adequately for another
series. Specification of proposal scales that produces fast-mixing kernels for
one data series may lead to slow mixing for another. In the following
experiment, we will use a single simulated time series, and choose schedules
that performs both well and poorly for this particular time series. The
objective is to see if the algorithm can recover from a relatively poorly
specified schedule and obtain reasonably accurate results.

\subsubsection{Results}

In this example we focus on the comparison between \smctwo and \pmcmc. We also
consider parallelized implementations of algorithms. In this case, due to its
relatively small number of chains, \pmcmc can be parallelized completely (and
often cannot fully utilize the hardware capability if a na\"\i ve approach to
parallelization is taken; while we appreciate that more sophisticated
parallelization strategies are possible, these depend instrinsicially upon the
model under investigation and the hardware employed and given our focus on
automatic and general algorithms, we don't consider such strategies here). The
\pmcmc algorithm under this setting is implemented such that each chain is
updated at each iteration. Further, for the \smc algorithms, we consider two
cases. In the first we can parallelize the algorithm completely (in the sense
that each core has a single particle associated with it). In this setting we
use a relatively small number of particles and a larger number of time steps.
In the second, we need a few passes to process a large number of particles at
each time step, and accordingly we use fewer time steps to maintain the same
total computation time. These two settings allow us to investigate the
trade-off between the number of particles and time steps. In both
implementations, we consider three schedules, $\alpha(t/T) = t/T$ (linear),
$\alpha(t/T) = (t/T)^5$ (prior), and $\alpha(t/T) = 1 - (1 - t/T)^5$
(posterior). In addition, the adaptive schedule based upon \cess is also
implemented for the \smctwo algorithm.

Results from 100 replicate runs of the two algorithms under various regimes
can be found in Tables~\ref{tab:pet-py-par-sel} and~\ref{tab:pet-bf-par-sel}
for the marginal likelihood and Bayes factor estimates, respectively. The \smc
algorithms consistently outperforms the \pmcmc algorithms in the parallel
settings. The Monte Carlo \sd of \smc algorithms is typically of the order of
one fifth of the corresponding estimates from \pmcmc in most scenarios. In
some settings with the smaller number of samples, the two algorithms can be
comparable.  Also at the lowest computational costs, the samplers with more
time steps and fewer particles outperform those with the converse
configuration by a fairly large margin in terms of estimator variance. It
shows that with limited resources, ensuring the similarity of consecutive
distributions, and thus good mixing, can be more beneficial than a larger
number of particles. However, when the computational budget is increased, the
difference becomes negligible. The robustness of \smc to the change of
schedules is again apparent.

It can also be seen that increasing the number of distributions not only
reduces the bias the path sampling estimator (as seen in the previous
example), but also reduces the variances considerably given the same number of
particles. On the other hand, increasing the number particles can only reduce
the variance of the estimates, in accordance with the central limit theorem;
see \citet{DelMoral:2006hc} for the standard estimator and extensions for
the path sampling estimator, Proposition~1{} in the supplementary material.
(as the bias arises from numerical integration approximation of the path
sampling estimator.)

\input{tab/pet-py-par-sel-zy}
\input{tab/pet-bf-par-sel-zy}

\paragraph{Effects of adaptive schedule}

A set of samplers with adaptive schedules are also used. Due to the nature of
the schedule, it cannot be controlled to have exactly the same number of time
steps as non-adaptive procedures. However, the \cess was controlled such that
the average number of time steps are comparable with the fixed schedules and
in most cases slightly less than the fixed numbers.

It is found that, with little computational overhead, adaptive schedules do
provide the best results (or very nearly so) and do so without user
intervention. The reduction of Monte Carlo \sd varies among different
configurations. For moderate or larger number of distributions, a reduction
about 50\% was observed. In addition, it shall be noted that, in this example,
the bias of path sampling estimates are much more sensitive to the
schedules than the previous Gaussian mixture model example. A vanilla linear
schedule does not provide a low bias estimator at all even when the number of
distributions is increased to a considerably larger number. The prior schedule
though provides a nearly unbiased estimator, there is no clear theoretical
evidence showing that this shall work for other situations. %Even it has more
%general usage, as suggested in \citet{Calderhead:2009bd}, the power still has
%to be chosen (in the previous \gmm example, $p = 2$ was the best choice while
%in this \pet example $p = 5$ is more suitable). In contrast,
The adaptive schedule, without any manual calibration, can provide a nearly
unbiased estimator, even when path-sampling is employed, in addition to
potential variance reduction.

\paragraph{Bias reduction for path sampling estimator}

As seen in Tables~\ref{tab:pet-py-par-sel} and~\ref{tab:pet-bf-par-sel}, a bad
choice of schedule $\alpha(t/T)$ can results in considerable bias for the
basic path sampling estimator, here for \smctwo-\ps but the problem is
independent of the mechanism by which the samples are obtained. Increasing the number of iterations can reduce
this bias but at the cost of additional computation time. As outlined in
Section~\ref{ssub:Improved Univariate Numerical Integration}, in the case of
the \smc algorithms discussed here, it is possible
to reduce the bias without increasing computational cost significantly. To
demonstrate the bias reduction effect, we constructed \smc sampler for the
above \pet example with only $1,000$ particles and about $20$ iterations
specified using the \cess based adaptive strategy. The path sampling estimator
was approximated using Equation~\eqref{eq:path_est} as well as other higher
order numerical integration or by integrating over a grid that contains
$\{\alpha_t\}$ at which the samples was generated. The results are shown in
Table~\ref{tab:pet-py-bias-reduction}

\input{tab/pet-bias}

\paragraph{Real data results}

Finally, the methodology of \smctwo-\ps was applied to measured positron
emission tomography data using the same compartmental setup as in the
simulations. The data shown in Figure~\ref{fig:petplot} comes from a study
into opioid receptor density in Epilepsy, with the data being described in
detail in \cite{Jiang:2009kf}. It is expected that there will be considerable
spatial smoothness to the estimates of the volume of distribution, as this is
in line with the biology of the system being somewhat regional. Some regions
will have much higher receptor density while others will be much lower,
yielding higher and lower values of the volume of distribution, respectively.
While we did not impose any spatial smoothness but rather estimated the
parameters independently for each time series at each spatial location, as can
be seen, smooth spatial estimates of the volume of distribution consistent
with neurological understanding were found using the approach. This method is
computationally feasible for the entire brain on a voxel-by-voxel basis, due
to the ease of parallelization of the \smc algorithm. In the analysis
performed here, 1000 particles were used, along with an adaptive schedule
using a constant $\cess^\star = 0.999$, resulting in about 180 to 200
intermediate distributions. The model selection results are very close to
those obtained by a previous study of the same data \citep{Zhou2013}, although
the present approach requires much less implementation effort and has roughly
the same computational cost.

\subsection{Summary}

These two illustrative applications and the \gmm example in the supplementary material have essentially shown three aspects of
using \smc as a generic tool for Bayesian model selection. First, as seen in
the \gmm example, all the different variants of \smc
proposed, including both direct and path sampling versions, produce results
which are competitive with other model selection methods such as \rjmcmc and
\pmcmc. In addition, in this somewhat simple example, \smctwo performs well,
and leads to low variance estimates with no appreciable bias. The effect of
adaptation was studied more carefully in the nonlinear \ode example, and it
was shown that using both adaptive selection of distributions as well as
adaptive proposal variances leads to very competitive algorithms, even against
those with significant manual tuning. This suggests that an automatic process
of model selection using \smctwo is possible. In the final example,
considering the easy parallelization of algorithms such as \smctwo suggests
that great gains in variance estimation can be made using settings such as
\gpu computing for application where computational resources are of particular
importance (such as in image analysis as in the PET example). It is also clear
that the negligible cost of the bias reduction techniques described means that
one should always consider using these to reduce the bias inherent in path
sampling estimation. As can also be seen in the supplementary material, there
is theoretical justification, in terms of a central limit theorem, available
for the path sampling estimator considered in \smctwo-\ps.

\section{Discussion}
\label{sec:Conclusion}

It has been shown that \smc is an effective Monte Carlo method for Bayesian
inference for the purpose of model comparison. Three approaches have been
outlined and investigated in several challenging scenarios.
The proposed strategy is always competitive and often substantially
outperforms the state of the art in this area.

Among the three approaches developed, \smcone is applicable to very general
settings. It can provide a robust alternative to \rjmcmc when inference on a
countable collection of models is required (and could be readily combined with
the approach of \cite{Jasra:2008bb} at the expense of a little additional
implementation effort). However, like all Monte Carlo methods involving
between model moves, it can be difficult to design efficient algorithms in
practice. The \smcthree algorithm is conceptually
appealing. However, specifying a suitable sequence of distributions between two
posterior distributions is challeging.

The \smctwo algorithm, which only involves within-model simulation, is most
straightforward to implement in many interesting problems and has been shown
to be exceedingly robust in many settings. As it depends largely upon a
collection of within-model \mcmc moves, any existing \mcmc algorithms can be
reused in the \smctwo framework. However, much less tuning is required because
the algorithm is fundamentally less sensitive to the mixing of the Markov
kernel and it is possible to implement effective adaptive strategies at little
computational cost. With adaptive placement of the intermediate distributions
and specification of the \mcmc kernel proposals, it provides a robust and
nearly automatic model comparison method.

Compared to the \pmcmc algorithm, \smctwo has greater flexibility in the
specification of distributions. Unlike \pmcmc, where the number and placement
of distributions can affect the mixing speed and hence performance
considerably, increasing the number of distributions will always benefit a
\smc sampler given the same number of particles. %When coupled with a path
%sampling estimator, this leads to less bias and variance; the \smc approach
%also provides an alternative estimator which avoids the numerical integration
%required in the path sampling estimator and hence alleviates any concerns over
%its contribution to the bias of the estimator.
Compared to its no-resampling
variant, it has been shown that \smc samplers with resampling can reduce the
variance of normalizing constant estimates considerably.

Even after three decades of intensive development, no Monte Carlo method can
solve the Bayesian model comparison problem completely automatically without
any manual tuning. However, \smc algorithms and the adaptive strategies
demonstrated in this paper show that even for realistic, interesting problems,
these samplers can provide good results with very minimal tuning and few
design difficulties. For many applications, they could already be used as near
automatic, robust solutions. For more challenging problems, they
can serve as solid foundation for the design of dedicated algorithms.

\section*{Supplementary Material}

Available from authors.

\small
\singlespace
\bibliographystyle{chicago}
\bibliography{paper}
%\todolist

\end{document}

%% file: macros.tex
\ifx\diff\undefined\DeclareMathOperator{\diff}{d}\fi
\ifx\var\undefined\DeclareMathOperator{\var}{var}\fi
\ifx\intd\undefined\newcommand{\intd}{\,\diff}\fi
\ifx\Exp \undefined\newcommand{\Exp} {\ensuremath{\mathbb{E}}\xspace}\fi
\ifx\Prob\undefined\newcommand{\Prob}{\ensuremath{\mathbb{P}}\xspace}\fi
\ifx\Real\undefined\newcommand{\Real}{\ensuremath{\mathbb{R}}\xspace}\fi

\newcommand{\rnd}[2]{\ensuremath{\frac{\diff#1}{\diff#2}}\xspace}

\newcommand{\data}{\ensuremath{\bm{y}}\xspace}
\newcommand{\mset}{\ensuremath{\mathcal{K}}\xspace}
\newcommand{\Mk}[1][k]{\ensuremath{{M_{#1}}}\xspace}
\newcommand{\paramk}[1][k]{\ensuremath{\theta_{#1}}\xspace}
\newcommand{\Paramk}[1][k]{\ensuremath{\Theta_{#1}}\xspace}

\newcommand{\rgamma}  {\ensuremath{\mathcal{G}}\xspace}

\newcommand{\bbI} {\ensuremath{\mathbb{I}}\xspace}

\newcommand{\calE}{\ensuremath{\mathcal{E}}\xspace}

\newcommand{\ess} {\ifmmode\text{ESS}\else ESS\xspace\fi}
\newcommand{\cess}{\ifmmode\text{CESS}\else CESS\xspace\fi}

\newcommand{\tick}{\checkmark}
\newcommand{\cross}{$\times$}

\def\ais     {AIS\xspace}
\def\bic     {BIC\xspace}
\def\mala    {MALA\xspace}
\def\mcmc    {MCMC\xspace}
\def\phm     {PHM\xspace}
\def\pmcmc   {PMCMC\xspace}
\def\rjmcmc  {RJMCMC\xspace}
\def\smc     {SMC\xspace}
\def\smcone  {SMC1\xspace}
\def\smcthree{SMC3\xspace}
\def\smctwo  {SMC2\xspace}

\def\ds{DS\xspace}
\def\ps{PS\xspace}
\def\sd{SD\xspace}

\def\gmm{GMM\xspace}
\def\ode{ODE\xspace}
\def\pet{PET\xspace}

\def\gpu{GPU\xspace}

\def\STATESKIP{\hskip.68cm}
\newcolumntype{C}{>{\centering\arraybackslash}X}

%% file: tab/node-s-all-ja.tex
\begin{table}
  \def\B{\color{blue}\it}
  \def\R{\color{red}\bf}
  \begingroup\small
    \begin{tabularx}{\linewidth}{lCClCCC}
      \toprule
      &&&& \multicolumn{2}{c}{Marginal likelihood} & \\
      &&&& \multicolumn{2}{c}{($\log p(\data|\Mk)\pm\text{\sd}$)} & \\
      \cmidrule(lr){5-6}
      $T$ & Proposal Scales & Annealing Scheme & Algorithm & $m = 3$ & $m = 5$ & Bayes factor $\log B_{3,5}$ \\ \midrule
      $10 $ & Manual    & Prior (5) & \pmcmc      & $-109.7\pm3.2$   & $-120.3\pm2.5$   & $10.6\pm3.8$ \\
      $30 $ &           &           &             & $\B-105.0\pm1.2$ & $\B-116.1\pm2.2$ & $\B11.2\pm2.5$ \\
      $100$ &           &           &             & $-134.7\pm7.9$   & $-144.1\pm6.2$   & $9.4\pm11.2$ \\ \midrule
      $500$ & Manual    & Prior (5) & \smctwo-\ds & $-104.6\pm2.0$   & $-112.7\pm1.8$   & $8.1\pm2.8$ \\
            &           &           & \smctwo-\ps & $-104.5\pm1.8$   & $-112.7\pm1.5$   & $8.2\pm2.5$ \\
      $500$ & Manual    & Adaptive  & \smctwo-\ds & $-104.5\pm1.1$   & $-112.7\pm1.1$   & $8.1\pm1.6$ \\
            &           &           & \smctwo-\ps & $-104.6\pm1.0$   & $-112.8\pm1.0$   & $8.2\pm1.5$ \\
      $500$ & Adaptive  & Adaptive  & \smctwo-\ds & $-104.5\pm0.5$   & $-112.7\pm0.4$   & $8.1\pm0.8$ \\
            &           &           & \smctwo-\ps & $\R-104.6\pm0.4$ & $\R-112.8\pm0.3$ & $\R8.1\pm0.6$ \\
      \bottomrule
    \end{tabularx}
  \endgroup
  \caption{Results for non-linear \ode models with data generated from simple
    model. {\B Italic}: Minimum variance for particular algorithm. {\R Bold}: Minimum variance among
    samplers.}
  \label{tab:node-s-all}
\end{table}

%% file: tab/node-c-all-ja.tex
\begin{table}
  \def\B{\color{blue}\it}
  \def\R{\color{red}\bf}
  \begingroup\small
    \begin{tabularx}{\linewidth}{lCClCCC}
      \toprule
      &&&& \multicolumn{2}{c}{Marginal likelihood} & \\
      &&&& \multicolumn{2}{c}{($\log p(\data|\Mk)\pm\text{\sd}$)} & \\
      \cmidrule(lr){5-6}
      $T$ & Proposal Scales & Annealing Scheme & Algorithm & $m = 3$ & $m = 5$ & Bayes factor $\log B_{5,3}$ \\ \midrule
      $10 $ & Manual    & Prior (5) & \pmcmc      & $-1651\pm27.9$   & $-85.1\pm36.6$   & $1566\pm42.1$ \\
      $30 $ &           &           &             & $\B-1640\pm7.4$  & $\B-78.9\pm11.2$ & $\B1561\pm12.8$ \\
      $100$ &           &           &             & $-1625\pm15.7$   & $-75.7\pm24.8$   & $1549\pm25.6$ \\ \midrule
      $500$ & Manual    & Prior (5) & \smctwo-\ds & $-1641\pm10.8$   & $-78.5\pm9.8$    & $1562\pm10.1$ \\
            &           &           & \smctwo-\ps & $-1641\pm 8.4$   & $-79.2\pm7.9$    & $1562\pm 8.5$ \\
      $500$ & Manual    & Adaptive  & \smctwo-\ds & $-1640\pm 6.9$   & $-78.6\pm4.8$    & $1561\pm7.1$ \\
            &           &           & \smctwo-\ps & $-1640\pm 5.4$   & $-78.8\pm3.7$    & $1561\pm6.8$ \\
      $500$ & Adaptive  & Adaptive  & \smctwo-\ds & $-1640\pm 2.2$   & $-79.4\pm1.7$    & $1560\pm3.1$ \\
            &           &           & \smctwo-\ps & $\R-1640\pm 1.9$ & $\R-78.5\pm1.5$  & $\R1562\pm2.3$ \\
      \bottomrule
    \end{tabularx}
  \endgroup
  \caption{Results for non-linear \ode models with data generated from complex
    model. {\B Italic}: Minimum variance for particular algorithm. {\R Bold}: Minimum variance among samplers.}
  \label{tab:node-c-all}
\end{table}

%% file: tab/pet-py-par-sel-zy.tex
\begin{table}
  \def\B{\color{blue}\it}
  \def\R{\color{red}\bf}
  \begingroup\small
    \begin{tabularx}{\linewidth}{lllCCCC}
      \toprule
      \multicolumn{3}{l}{Proposal scales}  & \multicolumn{3}{c}{Manual} & Adaptive \\
      \cmidrule(lr){1-3}\cmidrule(lr){4-6}\cmidrule(lr){7-7}
      \multicolumn{3}{l}{Annealing scheme} & Prior (5) & Posterior (5) & \multicolumn{2}{c}{Adaptive} \\
      \cmidrule(lr){1-3}\cmidrule(lr){4-4}\cmidrule(lr){5-5}\cmidrule(lr){6-7}
      $T$    & $N$   & Algorithm   & \multicolumn{4}{c}{Marginal likelihood estimates ($\log p(\data|\Mk)\pm\text{\sd}$)} \\ \midrule
      $500$  & $30 $ & \pmcmc      & $ -39.1\pm  0.56$ & $-926.8\pm376.99$ && \\
      $500$  & $192$ & \smctwo-\ds & $\B-39.2\pm0.25$ & $\B-39.7\pm1.06$ & $\B-39.2\pm0.18$ & $\R-39.1\pm0.12$ \\
             &       & \smctwo-\ps & $\B-39.2\pm0.25$ & $-91.3\pm21.69$  & $\B-39.2\pm0.18$ & $-39.1\pm0.13$   \\
      $100 $ & $960$ & \smctwo-\ds & $-39.3\pm0.36$   & $-40.6\pm1.41$   & $-39.2\pm0.31$   & $-39.2\pm0.19$   \\
             &       & \smctwo-\ps & $-39.3\pm0.35$   & $302.1\pm46.29$  & $-39.3\pm0.31$   & $-39.2\pm0.18$   \\ \midrule
      $5000$ & $30$  & \pmcmc      & $ -39.3\pm  0.21$ & $-917.6\pm129.54$ && \\
      $5000$ & $192$ & \smctwo-\ds & $-39.2\pm0.09$   & $\B-39.2\pm0.20$ & $-39.2\pm0.08$   & $-39.1\pm0.04$   \\
             &       & \smctwo-\ps & $-39.2\pm0.09$   & $-43.8\pm2.13$   & $-39.2\pm0.08$   & $-39.1\pm0.04$   \\
      $1000$ & $960$ & \smctwo-\ds & $\B-39.2\pm0.08$ & $-39.2\pm0.31$   & $\B-39.2\pm0.07$ & $\R-39.2\pm0.03$ \\
             &       & \smctwo-\ps & $\B-39.2\pm0.08$ & $-65.7\pm5.54$   & $\B-39.2\pm0.07$ & $\R-39.2\pm0.03$ \\
      \bottomrule
    \end{tabularx}
  \endgroup
  \caption{Marginal likelihood estimates of two component \pet model. $T$:
    Number of distributions in \smc and number of iterations used for
    inference in \pmcmc. $N$: Number of particles in \smc and number chains in
    \pmcmc. The \pmcmc and \smc with $N = 192$ are completely $N$-way
    parallelized.  \smc with $N = 960$ are $N/5$-way parallelized. {\B
      Italic}: Minimum variance for the same computational cost and the same proposal
    scales and annealing schemes.  {\R Bold}: Minimum variance for the same computaitonal
    cost and all proposal scales and annealing schemes.}
  \label{tab:pet-py-par-sel}
\end{table}

%% file: tab/pet-bf-par-sel-zy.tex
\begin{table}
  \def\B{\color{blue}\it}
  \def\R{\color{red}\bf}
  \begingroup\small
    \begin{tabularx}{\linewidth}{lllCCCC}
      \toprule
      \multicolumn{3}{l}{Proposal scales}  & \multicolumn{3}{c}{Manual} & Adaptive \\
      \cmidrule(lr){1-3}\cmidrule(lr){4-6}\cmidrule(lr){7-7}
      \multicolumn{3}{l}{Annealing scheme} & Prior (5) & Posterior (5) & \multicolumn{2}{c}{Adaptive} \\
      \cmidrule(lr){1-3}\cmidrule(lr){4-4}\cmidrule(lr){5-5}\cmidrule(lr){6-7}
      $T$    & $N$   & Algorithm   & \multicolumn{4}{c}{Bayes factor estimates ($\log B_{2,1}\pm\text{\sd}$)} \\ \midrule
      $500$  & $30 $ & \pmcmc      & $1.7\pm0.62$ & $-70.9\pm525.79$ && \\
      $500$  & $192$ & \smctwo-\ds & $\B1.6\pm0.27$ & $\B1.3\pm1.13$  & $\B1.6\pm0.20$ & $\R1.6\pm0.15$  \\
             &       & \smctwo-\ps & $\B1.6\pm0.27$ & $-3.9\pm30.02$  & $\B1.6\pm0.20$ & $\R1.6\pm0.15$  \\
      $100 $ & $960$ & \smctwo-\ds & $1.6\pm0.37$   & $0.5\pm1.55$    & $1.6\pm0.34$   & $1.6\pm0.21$    \\
             &       & \smctwo-\ps & $1.6\pm0.37$   & $-13.1\pm66.30$ & $1.6\pm0.33$   & $1.6\pm0.21$    \\ \midrule
      $5000$ & $30$  & \pmcmc      & $1.6\pm0.24$ & $-60.3\pm198.10$ && \\
      $5000$ & $192$ & \smctwo-\ds & $1.6\pm0.10$   & $\B1.6\pm0.23$  & $1.6\pm0.09$   & $1.6\pm0.05$    \\
             &       & \smctwo-\ps & $1.6\pm0.10$   & $1.3\pm2.98$    & $1.6\pm0.09$   & $1.6\pm0.05$    \\
      $1000$ & $960$ & \smctwo-\ds & $\B1.6\pm0.09$ & $1.6\pm0.33$    & $\B1.6\pm0.08$ & $\R1.6\pm0.04$  \\
             &       & \smctwo-\ps & $\B1.6\pm0.09$ & $-0.2\pm6.63$   & $\B1.6\pm0.08$ & $\R1.6\pm0.04$  \\
      \bottomrule
    \end{tabularx}
  \endgroup
  \caption{Bayes factor $B_{2,1}$ estimates of two component \pet model. $T$:
    Number of distributions in \smc and number of iterations used for
    inference in \pmcmc. $N$: Number of particles in \smc and number chains in
    \pmcmc. The \pmcmc and \smc with $N = 192$ are completely $N$-way
    parallelized.  \smc with $N = 960$ are $N/5$-way parallelized. {\B
      Italic}: Minimum variance for the same computational cost and the same schedule.
    {\R Bold}: Minimum variance for the same computational cost and all schedules.}
  \label{tab:pet-bf-par-sel}
\end{table}

%% file: tab/pet-bias.tex
\begin{table}\small
  \begin{tabularx}{\linewidth}{lXXXX}
    \toprule
    & \multicolumn{4}{c}{Number of grid points (compared to sampled iterations)} \\
    \cmidrule(lr){2-5}
    Integration rule & $\times1$ & $\times2$ & $\times4$ & $\times8$ \\
    \midrule
    Trapezoid
    & $-52.2\pm5.01$ & $-45.5\pm1.93$ & $-42.1\pm1.21$ & $-40.5\pm1.06$ \\
    Simpson
    & $-43.2\pm1.39$ & $-41.0\pm1.10$ & $-40.0\pm1.04$ & $-39.4\pm1.04$ \\
    Simpson $3/8$
    & $-42.1\pm1.21$ & $-40.5\pm1.06$ & $-39.7\pm1.04$ & $-39.3\pm1.04$ \\
    Boole
    & $-40.9\pm1.09$ & $-39.9\pm1.04$ & $-39.4\pm1.04$ & $-39.2\pm1.05$ \\
    \bottomrule
  \end{tabularx}
  \caption{Path sampling estimator of marginal likelihood of two component
    \pet model. The estimator was approximated using samples from \smctwo
    algorithm with $1,000$ particles and $20$ iterations, with different
    numerical integration strategies. Large sample result (see
    Table~\ref{tab:pet-py-par-sel}) provide an estimate of
    $-39.2$.}
  \label{tab:pet-py-bias-reduction}
\end{table}